\documentclass{article}
\usepackage{emulateapj}
\usepackage{apjfonts}

\newcommand{\proptwid}{\mathrel{\raise.3ex\hbox{$\propto$}\mkern-14mu
             \lower0.6ex\hbox{$\sim$}}}

\makeatletter
\let\@internalcite\cite
\def\cite{\def\astroncite##1##2{##1\ ##2}\@internalcite}
\def\citey{\def\astroncite##1##2{##1\ (##2)}\@internalcite}
\def\@citex[#1]#2{\if@filesw\immediate\write\@auxout{\string\citation{#2}}\fi
  \def\@citea{}\@cite{\@for\@citeb:=#2\do
    {\@citea\def\@citea{; }\@ifundefined
       {b@\@citeb}{{\bf ??}\@warning
       {Citation `\@citeb' on page \thepage \space undefined}}%
{\csname b@\@citeb\endcsname}}}{#1}}
\def\@cite#1#2{#1\if@tempswa #2\fi}
\def\@biblabel#1{}
\def\astroncite#1#2{#1\ #2}
\makeatother

\def\aproxgt{\mathrel{%
      \rlap{\raise 0.511ex \hbox{$>$}}{\lower 0.511ex \hbox{$\sim$}}}}
\def\aproxlt{\mathrel{%
      \rlap{\raise 0.511ex \hbox{$<$}}{\lower 0.511ex \hbox{$\sim$}}}}

\newcommand{\pfrac}[2]{\left(\frac{#1}{#2}\right)}

\newcommand{\A}{{\cal A}}
\newcommand{\D}{{\cal D}}
\newcommand{\G}{\Gamma}
\newcommand{\g}{\gamma}
\newcommand{\Gt}{\tilde{\Gamma}}
\newcommand{\Go}{\G_0}
\newcommand{\Gf}{\G_f}
\newcommand{\brk}{{\rm break}}
\newcommand{\peak}{{\rm peak}}
\newcommand{\Min}{{\rm min}}
\newcommand{\Max}{{\rm max}}
\newcommand{\e}{\varepsilon}
\newcommand{\crit}{{\rm crit}}
\newcommand{\obs}{{\rm obs}}
\newcommand{\ext}{{\rm ext}}
\newcommand{\dyn}{{\rm dyn}}
\newcommand{\syn}{{\rm syn}}

\def\be{\begin{equation}}
\def\ee{\end{equation}}
\def\bea{\begin{eqnarray}}
\def\eea{\end{eqnarray}}

\begin{document}

\slugcomment{Submitted to The Astrophysical Journal, October 13, 1998}

\lefthead{CHIANG}
\righthead{TIME-INTEGRATED GAMMA-RAY BURST SYNCHROTRON SPECTRA}

\title{Time-Integrated Gamma-Ray Burst Synchrotron Spectra from 
Blast Wave/Cloud Interactions}

\author{James Chiang\altaffilmark{1}}

\affil{E. O. Hulburt Center for Space Research, Code 7653, Naval Research 
Laboratory, Washington DC 20375-5352}
\altaffiltext{1}{Present Address: JILA, Campus Box 440,
University of Colorado, Boulder CO 90309-0440}

\begin{abstract}
We show that the spectral shape of the low energy tails found for the
time-integrated spectra of gamma-ray bursts, even in the absence of
strong synchrotron cooling, can be significantly softer than the $\nu
F_\nu \propto \nu^{4/3}$ asymptote predicted by synchrotron shock
models.  As we have noted in a previous work, blast wave deceleration
via interaction with ambient material causes the characteristic
electron injection energy to decrease in proportion to the bulk
Lorentz factor of the blast wave, and under certain conditions, this
effect will at least partially account for the observed increase in
pulse widths with decreasing energy.  This spectral softening can also
be reflected in the time-integrated pulse spectrum.  Using a simple
model for the blast wave interaction with a dense cloud of material,
we show that just below the $\nu F_\nu$ spectral peak the integrated
spectrum behaves as $\nu F_\nu \sim \nu^{1/2}$ and rolls over to a
$\nu^{4/3}$ dependence at lower energies, thus a spectral shape arises
which is similar to that predicted for the spectrum of a strongly
synchrotron-cooled electron population.  We discuss the implications
of this work in the context of models of burst light curve variability
which are based on blast wave/cloud interactions.
\end{abstract}


\section{Introduction}
Models of blast wave shells interacting with ambient density
inhomogeneities have been proposed as a possible source of the
variability seen in GRB light curves (e.g., Sari \& Piran 1995).
However, external shock models of this type have come under criticism,
either because of the enormous blast wave energies which are required,
$\ga 10^{53}$~ergs (Sari \& Piran 1997), or because of difficulties in
producing the observed temporal structure (Fenimore, Madras, \&
Nayakshin 1996; Fenimore et al.\ 1998).  With the measurement of the
redshift and afterglow flux for GRB~971214 (Kulkarni et al.\ 1998), it
is now clear that isotropic blast wave energies of $\ga 10^{53}$~ergs
do occur in some bursts.  This result has motivated a more recent
treatment by Dermer \& Mitman (1998) of blast wave/cloud interactions
and the production of complex burst light curves.  In that work, the
overall burst light curve structure was modeled, but the detailed
spectral and temporal shape of individual pulses was not.  We consider
some aspects of these latter issues in this paper.

Several authors have analyzed the spectra of the prompt high energy
emission from gamma-ray bursts (GRBs).  In particular, data from the
Burst and Transient Source Experiment (BATSE) aboard the {\em Compton
Observatory} have provided detailed spectral information which
continues to constrain models of emission from these objects.  Among
others, Katz (1994) and Tavani (1996) have argued persuasively that
the spectral shape of the prompt emission from most bursts is due to
synchrotron radiation produced by electrons accelerated by a
relativistic shock.  Such electrons will have a well-defined
characteristic energy, and Katz (1994) has pointed out that in the
absence of energy losses the instantaneous low energy tail of the
synchrotron spectrum from these electrons will have spectral
dependence $\nu F_\nu \propto \nu^\delta$ where the index $\delta =
4/3$ (Rybicki \& Lightman 1979).  Radiative or adiabatic losses will
cause the electron energies to evolve so that the instantaneous
spectral index will satisfy $\delta \le 4/3$.  If synchrotron cooling
is the dominant energy loss mechanism, then just below the
characteristic synchrotron photon energy the spectrum will have
$\delta = 1/2$ (Cohen et al.\ 1997; Sari, Piran, \& Narayan 1998) .

From the examination of BATSE data, two recent studies indeed find
that the above constraint appears to be met by a large majority of
bursts.  Cohen et al.\ (1997) find that the low energy spectral
indices of all 11 GRBs they examined satisfy $\delta \le 4/3$ and that
for several bursts the index approaches $4/3$ at the lowest energies.
Schaefer et al.\ (1998) consider spectra from over 100 bursts from
BATSE and Ginga data and find that 90\% of them appear to satisfy this
limit as well.  Of the remaining 10\%, for only one burst can no
mitigating complication, such as background uncertainties or data
gaps, be found.  For much of the energy range below the $\nu F_\nu$
peak, most of the spectral indices found by both groups lie at
intermediate values well within the range $1/2 < \delta < 4/3$.
Schaefer et al.\ form a composite spectrum of 19 bright BATSE bursts
and provide a quantitative estimate of the characteristic index.  They
find that the low energy index of the composite spectrum is $\delta
\simeq 0.77$.  Cohen et al.\ suggest that intermediate values of
$\delta$ result from the heterogeneous nature of the emitting
material, presumably with some regions being strongly cooled and
others less so.  However, in a more recent study of a somewhat larger
BATSE burst sample, Preece et al.\ (1998) claim that as many as 25\%
of the bursts studied violate the $\delta < 4/3$ limit.  This implies
that synchrotron radiation cannot be the source of the gamma-ray
emission for these bursts.  However, Preece also has noted that
violations of this limit can depend upon the specific spectral model
chosen to fit the data (see discussion in Schaefer et al.\ 1998).

In this paper we will concern ourselves exclusively with synchrotron
radiation and processes which can soften the apparent low energy
spectrum.  In a previous paper (Chiang 1998), we showed how the
reduction of the characteristic electron injection energy due to the
deceleration of the blast wave results in a spectral softening such
that burst pulses are broader at lower energies.  Fenimore et al.\
(1995) have noted this phenomenon in BATSE burst data, and several
authors have attributed it to synchrotron cooling (Kazanas, Titarchuk,
\& Hua 1998; Dermer 1998).  In the present work, we show that the
time-integrated spectrum of a relativistic blast wave shell
interacting with a dense cloud of material naturally produces a low
energy spectral index $\delta \simeq 1/2$ which rolls over to $4/3$ at
lower energies.  Again, this effect is due to blast wave deceleration
and the corresponding reduction of the characteristic electron
injection energy even in the absence of synchrotron cooling.

In the remainder of this paper, we present in \S2 our model for the
deceleration of the blast wave as it interacts with a dense cloud; in
\S3, we describe the evolution of the electron distribution function
in the non-radiative limit and compute instantaneous and
time-integrated spectra as well as pulse profiles at various energies;
and finally, in \S4, we discuss this work's implications for and
applicability to actual burst models.

\section{Blast Wave/Cloud Interactions in the Non-Radiative Limit}

Although blast waves are generally treated as spherical, for the
interaction of a blast wave shell with a cloud at large radius, $R$,
we consider the portion of the shell which intercepts a sufficiently
small cloud to be planar.  This implies $l \ll R$, where $l$ is the
linear size of the cloud perpendicular to the trajectory of the blast
wave.  For the purposes of this paper, we assume that the blast wave
shell has traveled through a vacuum from $R=0$ until it encounters the
cloud.  It is therefore initially thin, cold and dense, the internal
energy of the material from the initial fireball event having been
converted to the bulk motion of the expanding material.  The initial
mass of the shell fragment intercepting the cloud is $M_0 = E_0
\A/4\pi R^2 c^2 \Go$ where $E_0$ is the fireball energy, $\Go$ is the
initial bulk Lorentz factor of the shell and $\A \sim l^2$ is the
cross-sectional area of the cloud which we take to be constant.  The
evolution of the bulk Lorentz factor as a function of the penetration
distance $r$ into the cloud is given by energy and momentum
conservation (cf.\ Chiang \& Dermer 1998):
\begin{eqnarray}
\G(r) &=& \frac{\Go M_0 + m(r)}{(M_0^2 + 2\Go M_0 m(r) + m(r)^2)^{1/2}}
\label{G(r)_exact}\\
      &\simeq& \frac{\Go M_0}{(2\Go M_0 m(r))^{1/2}},
\label{G(r)_approx}
\end{eqnarray}
where $m(r) = \A n_\ext m_p r$ is the cloud rest mass which has been
swept-up as a function of $r$, and $n_\ext$ is the cloud number
density.  The approximate expression (eq.~\ref{G(r)_approx}) is valid
during the deceleration phase but while the shell Lorentz factor is
still relativistic, i.e., for $\Go > \G \gg 1$.  The deceleration
phase occurs beyond a penetration depth $r_d = M_0/\A n_\ext m_p \Go$.
We consider clouds which are sufficiently dense so that $r_d \ll r_0$
where $r_0$ is the size of the cloud along the trajectory of the blast
wave shell.

Equation~\ref{G(r)_exact} corresponds to the non-radiative limit.
This limit in turn places a constraint on the strength of the magnetic
field generated during the shell/cloud interaction in which the
synchrotron cooling timescale must exceed the dynamical timescale in
the co-moving frame of the shell.  We define the dynamical timescale
as the time for the shell to cross the cloud.  From
eq.~\ref{G(r)_approx}, we find $\tau_\dyn \simeq (r_0/c)(m(r_0)/\Go
M_0)^{1/2}$, and using the familiar expressions for the synchrotron
loss timescales we find $\tau_\syn \simeq (m_e/m_p)^2(c \sigma_T
\zeta_c n_\ext \xi_e \xi_B)^{-1} \Gt^{-3}$.  Here $\zeta_c \sim 4$--7
is the compression ratio of the shock in the instantaneously co-moving
frame, $\Gt$ is a characteristic bulk Lorentz factor over the
traversal time of the cloud, and $\xi_e$ and $\xi_B$ are equipartition
parameters for the electrons and magnetic field respectively.  These
quantities are defined in the usual way: the characteristic electron
injection Lorentz factor is $\g_e = \xi_e (m_p/m_e) \G$, and the
magnetic field energy density is given by $B^2/8\pi = \xi_B \zeta_c
n_\ext m_p c^2 \G^2$ (Chiang \& Dermer 1998).  The resulting
non-radiative constraint is then
\begin{equation}
\xi_e \xi_B \la 18\pi \frac{m_p c^2}{\sigma_T} \pfrac{m_e}{m_p}^2  
                \frac{1}{\zeta_c} \frac{R^2}{E_0}\pfrac{\Gf}{\Gt}^3
\label{xi_limit}
\end{equation}
where $\Gf$ is the bulk Lorentz factor at $r = r_0$.  Assuming $\Go =
200$, we take a reasonable value to be $\Gf=20$ so that most of the
bulk kinetic energy of the shell is converted to internal energy, but
it nonetheless remains relativistic, and our approximations (e.g.,
eq.~\ref{G(r)_approx}) still apply.  Note that this expression is
independent of the cloud density and size.  For $E_0 = 10^{54}\,$ergs,
$R = 10^{16}\,$cm, we have $\xi_e \xi_b \la 10^{-6} (\Gf/\Gt)^3$.
This limit is similar to that for non-radiative evolution found by
Chiang \& Dermer~(1998) for a spherical blast wave which interacts
with a uniform external medium.  More detailed calculations show that
this limit is actually conservative.  For $\Gt = \sqrt{\Go\Gf} =
\sqrt{10}\Gf$, we can set $\xi_e = 10^{-1}$ and $\xi_B = 10^{-6}$ and
will do so unless otherwise noted.

An appropriate cloud size can be inferred from pulse widths seen in
actual burst data.  Norris et al.\ (1996) examined the distribution of
individual pulse widths for a large number of burst light curves.
They found that widths range from $10^{-2}$ to $10\,$s.  In the hope
of applying the signal from the interaction we describe to individual
pulses seen in actual burst data, we choose $\Delta t_{\rm pulse} =
0.1\,$s as an exemplary value.  We will equate this to the traversal
time for the blast wave to cross the cloud as inferred by a distant
observer lying along the shell fragment trajectory.  As we will see
below, this will give an upper limit to the actual pulse width.  We
find $\Delta t_{\rm pulse} \simeq r_0/4 c \Gf$.  This yields $r_0 \la
10^{13}\,$cm.  Solving for the accumulated cloud rest mass as a
function of $\G$, we obtain $m \simeq M_0\Go/2\G^2$ and find a cloud
density $n_\ext \ga 10^7\,$cm$^{-3}$.  We note that for an initial
bulk Lorentz factor of $\Go = 200$ and if we assume $l \sim r_0$ we
have $l/R < 1/\Go$, so that we can largely ignore curvature effects in
computing the pulse profiles (Fenimore et al.\ 1998).

\section{Results}

\subsection{Instantaneous and Time-Integrated Spectra}

We now compute the electron distribution function in the co-moving
frame of the shell. Since we are only concerned with the low energy
tail of the synchrotron spectrum, we model the electron injection
function at any given depth $r$ to be a $\delta$-function in energy:
$dN/drd\g = \A n_\ext \delta(\g - \g_e(r))$ where $\g_e(r) = \xi_e
(m_p/m_e)\G(r)$.  Evaluating the $\delta$-function and integrating
over $r$, we find
\begin{eqnarray}
\frac{dN}{d\g} &\simeq& \frac{M_0 \Go}{m_p} \frac{1}{\xi_e}
                        \frac{m_e}{m_p} \G^{-3}\\
               &\simeq& \frac{M_0 \Go}{m_p}\left(\frac{1}{\xi_e}
                        \frac{m_e}{m_p}\right)^2\g^{-3}.
\end{eqnarray}

The instantaneous synchrotron spectrum from such an electron
distribution will be flat with $\nu L_\nu \propto \nu^0$.  As seen by
an observer along the shell trajectory, this flat portion of the
spectrum will obtain for energies $\e \equiv h\nu/m_e c^2$ bounded by
\begin{eqnarray}
\e_\Max &=& 2\left(\xi_e \frac{m_p}{m_e}\right)^2 \pfrac{B_\infty}{B_\crit}
                \G_0^2 \G^2
\label{emax}                \\
\e_\Min &=& 2\left(\xi_e \frac{m_p}{m_e}\right)^2 \pfrac{B_\infty}{B_\crit}
                \G^4,
\label{emin}
\end{eqnarray}
where $B_\infty \equiv (8\pi \zeta_c n_\ext m_p c^2 \xi_B)^{1/2}$ and
$B_\crit = m_e^2 c^3/e \hbar = 4.4 \times 10^{13}\,$G, and we have
estimated the Doppler factor for the shell to be $\D \simeq 2\G$,
appropriate for an observing angle of zero.  For $\e < \e_\Min$ the
spectrum will behave as $\nu L_\nu \propto \e^{4/3}$ while above
$\e_\Max$ the spectrum will be sharply cut-off.  The normalization of
this spectrum can be estimated using the $\delta$-function
approximation for synchrotron radiation (Dermer, Sturner, \&
Schlickeiser 1997):
\begin{eqnarray}
\nu L_\nu &\simeq& \frac{2 c \sigma_T}{3}\frac{B^2}{8\pi}
                   \pfrac{B_\crit}{B}^{3/2} \pfrac{\e}{\D}^{3/2}
                   \frac{dN}{d\g} \D^4\\ 
          &\simeq& \frac{32}{3} c^3
                   \sigma_T \pfrac{m_p}{m_e}^2 \zeta_c n_\ext \xi_B
                   \xi_e^2 M_0 \Go \G^6.
\label{nuLnu}
\end{eqnarray}
Here we note that for a $\nu L_\nu$ spectrum which is isotropic in the
co-moving frame the amplification due to Doppler boosting is $\D^4$.

Defining the observer time $t_\obs$ to begin at $r=r_d$, we use
eq.~\ref{G(r)_approx} and find the time dependence of the bulk Lorentz
factor to be $\G \simeq (M_0 \Go/8 c m_p n_\ext \A t_\obs)^{1/4}$.
Hence, we have $\nu L_\nu \propto t_\obs^{-3/2}$, $\e_\Max \propto
t_\obs^{-1/2}$ and $\e_\Min \propto t_\obs^{-1}$.  Therefore, for a
given energy, the flat portion of the instantaneous $\nu L_\nu$
spectrum enters into a narrow observed band pass at energy $\e$ at a
time $t_\Min \propto \e^{-1}$ and exits at a later time $t_\Max
\propto \e^{-2}$.  The observed energy dependence of the
time-integrated spectrum will then be given by
\begin{eqnarray}
\left.\frac{dE}{dA}\right|_\e
    &\propto& \int_{t_\Min(\e)}^{t_\Max(\e)} t_\obs^{-3/2} dt_\obs =
              2(t_\Min^{-1/2} - t_\Max^{-1/2})\\
    &\sim& \e^{1/2}.
\label{e_dependence}
\end{eqnarray}
The lower limit provides the dominant term and thus gives the
spectral index $\delta = 1/2$.  This energy dependence will extend
from a maximum energy 
\begin{equation}
\e_\peak \simeq 2\left(\xi_e \frac{m_p}{m_e}\right)^2 
                   \pfrac{B_\infty}{B_\crit}\Go^4
\label{e_peak}
\end{equation}
 to a break energy
\begin{equation}
\e_\brk \simeq 2\left(\xi_e\frac{m_p}{m_e}\right)^2\pfrac{B_\infty}{B_\crit}
            \pfrac{E_0}{8\pi m_p c^2 r_0 n_\ext R^2}^2.
\label{e_break}
\end{equation}
This is the characteristic photon energy from the electrons injected
at $r_0$.  At this break energy, we use eq.~\ref{nuLnu} and estimate
the spectral fluence explicitly:
\begin{equation}
\left.\frac{dE}{dA}\right|_{\e=\e_\brk} 
          \simeq \frac{1}{3 \pi d_l^2} \frac{\sigma_T}{m_e^2 c^4}
                 \frac{\zeta_c \A}{n_\ext r_0} \xi_B \xi_e^2 
                 \pfrac{E_0}{4 \pi R^2}^3,
\label{break_fluence}
\end{equation}
where $d_l$ is the luminosity distance to the source.  We have
neglected factors of $(1+z)$ for the sake of brevity.  Below the break
energy, the spectrum will consist of the low energy synchrotron tail
and behave as $\sim \e^{4/3}$.  We also estimate the spectral fluence
at the peak:
\begin{equation}
\left.\frac{dE}{dA}\right|_{\e=\e_\peak} 
          \simeq \frac{2}{3 \pi d_l^2} \frac{m_p \sigma_T}{m_e^2 c^2}
               \zeta_c \A \xi_B \xi_e^2 \pfrac{E_0}{4\pi R^2}^2 \Go^2.
\label{peak_fluence}
\end{equation}

\noindent
\parbox{0.45\textwidth}{
\centerline{\epsfysize=0.45\textwidth\epsfbox{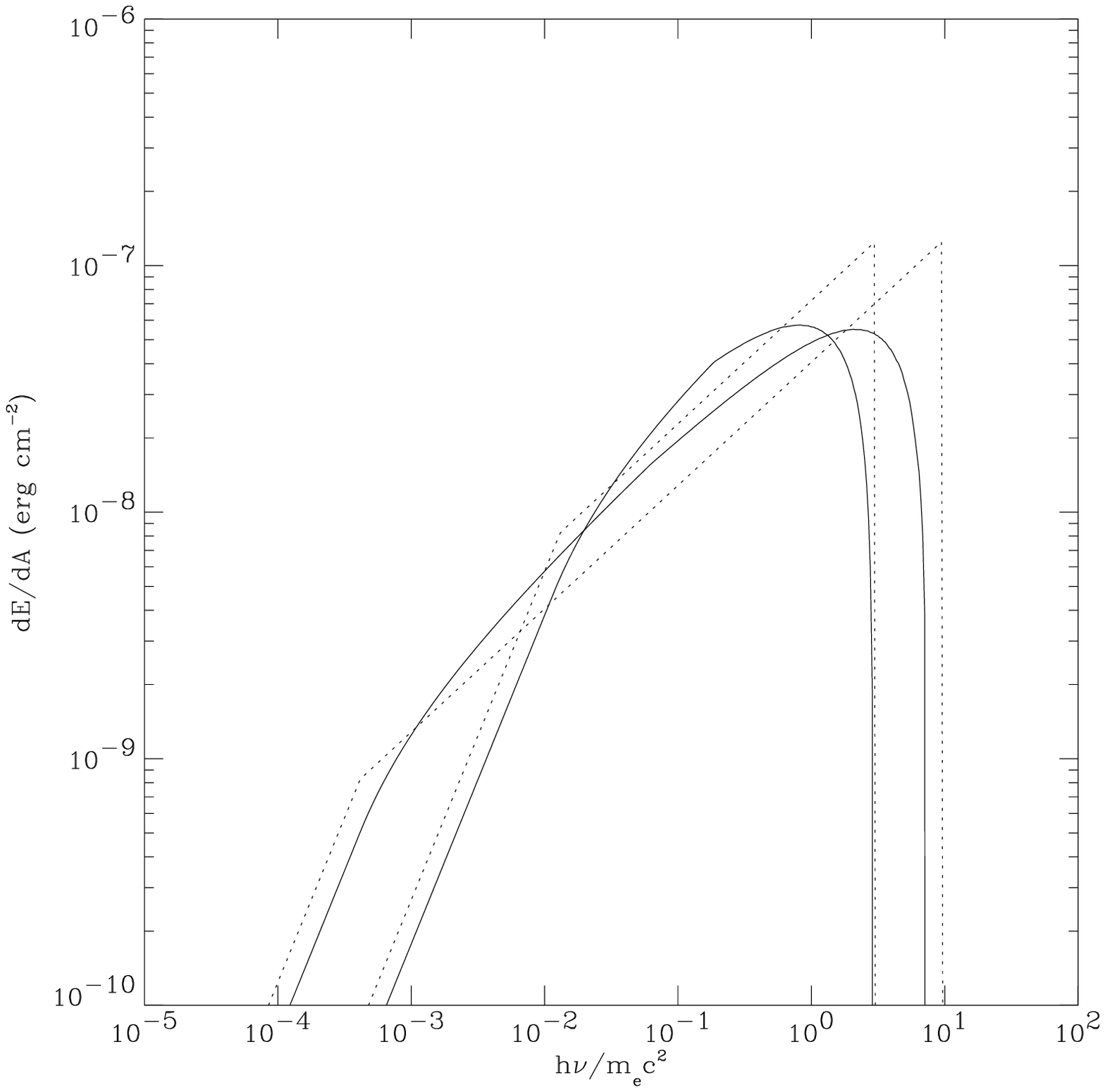}}
\figcaption{\small Time-integrated spectra for a blast-wave shell 
interacting with a small dense cloud.  The solid curves are the
numerical integrations and the dotted curves are the analytic
estimates. Model parameters are $\Go=200$, $E_0 = 10^{54}$~ergs, $R =
10^{16}$~cm, $r_0 = l = 10^{13}$~cm, $\xi_e = 10^{-1}$, $\xi_B =
10^{-6}$, $d_l = 1.6 \times 10^{28}$~cm.  The cloud density for the
spectrum with the higher break energy is $n_\ext = 10^7$~cm$^{-3}$,
while for the spectrum with the lower break energy $n_\ext =
10^8$~cm$^{-3}$ (see eq.~\protect{\ref{e_break}}).
\label{cb_spectra}}}
\medskip

In Figure~\ref{cb_spectra} we plot the time-integrated spectra for a
blast wave interacting with clouds of two different densities, $n_\ext
= 10^7$ and $10^8$~cm$^{-3}$, keeping the other model parameters the
same in both cases.  The solid curves are spectra for which we have
used the full expressions for the bulk Lorentz factor
(eq.~\ref{G(r)_exact}),
the electron distribution function and the energy bounds for the flat
part of the $\nu L_\nu$ spectrum.  The only numerical approximations
we used are the $\delta$-function expression for the synchrotron
emissivity and the assumption that below $\e_\Min$ the instantaneous
spectrum has a $\e^{4/3}$ dependence.  The dotted curves are our
analytic estimates (eqs.~\ref{e_dependence}--\ref{peak_fluence}) for
the integrated spectra.

From these spectra and from eq.~\ref{e_break} we see that the location
of the roll-over to the $\e^{4/3}$ spectral dependence is a sensitive
function of several important burst parameters---namely, the isotropic
fireball energy $E_0$, the cloud size $r_0$, the cloud density
$n_\ext$ and the radial location of the cloud $R$.  The quantity $E_0$
can be estimated by integrating up the total prompt burst and
afterglow flux (e.g., Kulkarni et al.\ 1998 for GRB~971214), while $R$
can be constrained via afterglow redshift and radio scintillation
measurements as was done for GRB~970508 (Metzger et al.\ 1997; Frail
et al.\ 1997).  Note also that the initial bulk Lorentz factor $\Go$
is strongly constrained by equating $\e_\peak$ with the energy of the
spectral peak.  This leaves the combination $n_\ext r_0$ which is the
column density through the cloud along the line-of-sight.  The burst
spectra depicted in Cohen et al.\ (1997), especially those of
GRB~920406, GRB~910601 (rise) and GRB~930201, place an upper limit of
$\sim 1$~decade on the range over which the $\nu^{1/2}$ dependence is
seen.  Equating this factor of 10 to the ratio $\e_\peak/\e_\brk$, we
obtain
\begin{equation}
n_\ext r_0 \la 10^{19} \pfrac{E_0}{10^{54}\mbox{erg}}
                       \pfrac{R}{10^{16}\mbox{cm}}^{-2}
                       \pfrac{\Go}{200}^{-2}\mbox{cm}^{-2}
\end{equation}
As we have shown above, individual pulse durations can then place
additional limits on $r_0$ and $n_\ext$, rendering all the physical
parameters of the blast wave/cloud interaction reasonably
well constrained.

In producing these fluence spectra, we have used a luminosity distance
of $d_l = 1.6 \times 10^{28}$~cm, corresponding to a redshift of $z =
1$ with $H_0 = 67\,$km~s$^{-1}\,$Mpc$^{-1}$ and $q_0 = 0.5$.  Thus
even though we have considered very weak magnetic fields so that the
electrons are not radiatively efficient, we see that for an individual
pulse, the received fluence is in accord with total burst fluences
which are typically seen.  Approximately 20 of these sub-0.1 second
pulses will be required to make up burst profiles with total fluences
of order $\sim 10^{-6}$~erg~cm$^{-2}$.

\subsection{Pulse Profiles}

At energies satisfying $\e_\Min(t_\obs) < \e < \e_\Max(t_\obs)$, the
temporal dependence of the light curve is $\sim t_\obs^{-3/2}$.  When
$\e = \e_\Min(t_\obs)$, there will be a break in the light curve.
This marks the transition from the $\e^{4/3}$ part of the
instantaneous synchrotron spectrum to the $\e^0$ part.  Inverting the
expression for $\e_\Min$, we estimate the temporal break to occur at
\begin{equation}
t_{\obs,\brk} \simeq 2\left(\xi_e \frac{m_p}{m_e}\right)^2
                    \pfrac{B_\infty}{B_\crit} \pfrac{M_0\Go}{8 c
                     \A n_\ext m_p} \e^{-1}.
\end{equation}
For observer times just prior to $t_{\obs,\brk}$, the light curve will
have a $t_\obs^{-1/6}$ dependence.  In Figure~\ref{cb_light_curves} we
plot light curves obtained from the numerical calculations presented
in Figure~\ref{cb_spectra}.  The energies of these light curves are
the lower bounds of the four standard BATSE LAD channels---25, 57, 115
and 320~keV (see Fenimore et al.\ 1995).  The solid curves are the
$n_\ext = 10^7$~cm$^{-2}$ calculation, while the dashed curves are the
$n_\ext = 10^8$~cm$^{-2}$ case.  Each light curve has been normalized
to have maximum intensity of unity.  The onset of the $t_\obs^{-3/2}$
decline is delayed at lower energies according to $t_{\obs,\brk}
\propto \e^{-1}$.  This energy dependence disagrees with the $\Delta t
\propto \e^{-0.45}$ result found by Fenimore et al.\ (1995).  However,
at early times, when the observed flux at all energies of interest is
due to the $\e^{4/3}$ portion of the spectra, the normalized light
curves all decline together.  So, in some average sense, the pulse
widths in this model may be able to accommodate the energy dependence
found by Fenimore et al..

\section{Implications for Burst Data and Models}

This model of GRB pulses produces a time-integrated spectral shape
which is very similar to that expected for the spectrum of a strongly
synchrotron-cooled electron distribution, even though synchrotron
losses themselves are negligible.  Therefore, spectral considerations
alone may not be able to determine the importance of strong
synchrotron cooling in prompt burst light curve variability.  In
particular, this model offers an alternative to the suggestion by
Cohen et al.\ (1997) that the rapid variability seen in the light
curve of GRB~930201 is due to short radiative timescales and that the
$\delta\simeq 1/2$ spectral index in the time-integrated spectrum is
corroborating evidence of this.  The present picture is that the sharp
spikes such as those in the GRB~930201 light curve are due to
non-radiative blast-wave interactions with dense clouds.  The short
timescale variability and the spectral shape arise because of the
quick deceleration of the blast wave.  The former is largely
controlled by the associated reduction in the amplification due to
Doppler boosting while the latter mostly results from the decreasing
characteristic electron injection energy.  As can be seen from
Figures~\ref{cb_spectra} \&~\ref{cb_light_curves}, larger cloud
densities result in shorter pulse timescales and also in softer
time-integrated spectra which is exactly the same correlation
attributed to more efficient synchrotron cooling.

\noindent
\parbox{0.45\textwidth}{
\centerline{\epsfysize=0.45\textwidth\epsfbox{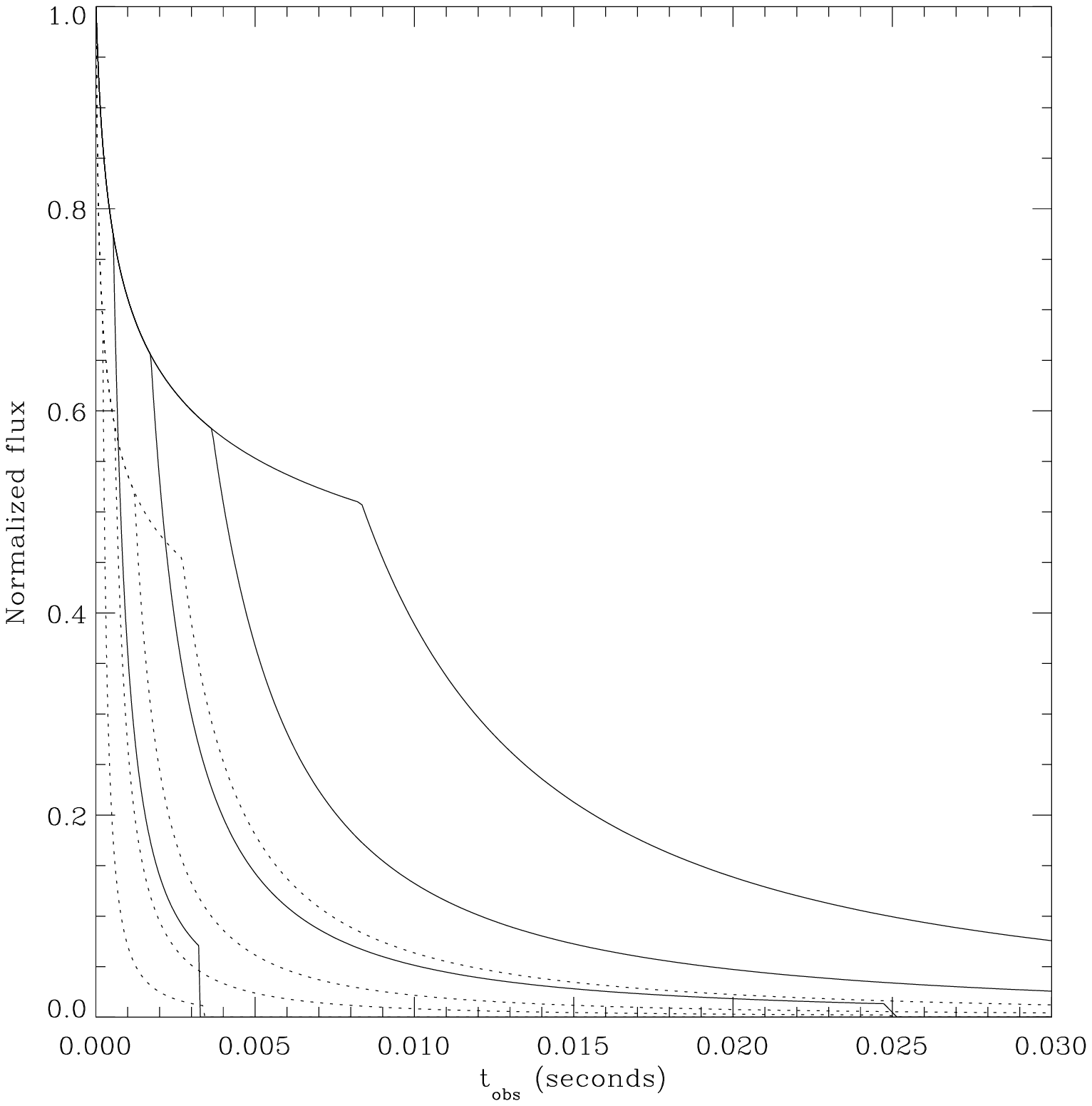}}
\figcaption{\small Blast wave/cloud burst pulse profiles at energies 
25, 57, 115 and 320~keV which are the lower bounds of the four
standard BATSE LAD channels.  The narrower light curves correspond to
the higher energies.  The solid curves are produced using $n_\ext =
10^7$~cm$^{-3}$ and the dotted curves are for $n_\ext =
10^8$~cm$^{-3}$.  All the other parameters are as given in
Figure~\protect{\ref{cb_spectra}}.
\label{cb_light_curves}}}
\medskip

If synchrotron cooling were somewhat efficient, then we expect that
the pulses would actually be broader in time.  This would occur if the
liberation of accumulated internal energy at later times acts to
counter the reduction in received flux due to the broadening Doppler
cone.  This argument runs counter to the suggestion by Cohen et al.\
(1997) that shorter pulse durations should always be the result of
shorter radiative timescales.  However, if the blast wave interaction
were {\em perfectly} radiatively efficient, then it does seem
reasonable that the pulse durations would be shorter than in the
non-radiative case we have examined here, not merely because of the
shorter radiative timescales, but also because the bulk Lorentz factor
would decline more rapidly.  We suggest that it is in the intermediate
case, where the dynamical timescale is a significant fraction of the
synchrotron loss timescale, that the pulse durations will be longest.

Certain assumptions of this model force the time-integrated spectral
index to be $\delta = 1/2$.  In a spherical, rather than planar
geometry, the time dependence of the bulk Lorentz factor during the
deceleration phase will be $\G \sim t_\obs^{-g/(2g+1)}$ where $g$ is
the power-law index of the bulk Lorentz factor radial dependence, $\G
\propto R^{-g}$, where $3/2 < g < 3$ (Chiang \& Dermer 1998; Dermer,
Chiang, \& B\"ottcher 1998).  For non-radiative evolution, $g = 3/2$.
Neglecting adiabatic losses, this implies a time dependence $\nu L_\nu
\propto t_\obs^{-9/4}$ or a time-integrated low energy spectral index
of $\delta = 5/4$ (cf. eqs.~\ref{nuLnu} \& \ref{e_dependence}).
Inclusion of adiabatic losses may soften the spectrum further, but we
note that the estimate by Cohen et al.\ (1997) of the effect of these
losses for an expanding blast wave geometry, but neglecting some of
the effects of deceleration considered here, gives exactly this same
spectral dependence.

Another important assumption is that the cloud density is a constant,
independent of the penetration depth $r$.  A density which decreases
with $r$ will leave the spectrum harder, while an increasing density
will likely make the spectrum even softer.  Likewise, a range of cloud
column densities will have the effect of producing an intermediate
spectral index for the composite spectrum integrated over many clouds.
Furthermore, we note that even a single cloud will have an apparent
index differing from $\delta = 1/2$ if the cloud column density is
sufficiently small so that the break energy is relatively large and
only the gradual roll-over to $\delta = 4/3$ is seen.  The $n_\ext =
10^7$~cm$^{-3}$ case shown in Figure~\ref{cb_spectra} is an example of
this effect.

This model has several features which might make it difficult to fit
into the context of realistic burst source models.  First of all,
clouds with densities of $\sim 10^7$~cm$^{-3}$ in an otherwise
evacuated surrounding environment may not be physically realizable.
In models of AGNs, the dense clouds which are posited to compose the
broad line region require a confinement mechanism such as a strong
external magnetic field or hot external ambient matter in order that
they not dissipate on the relevant timescales.  If confinement is
necessary for dense clouds around burst progenitors, it is not clear
that either of these mechanisms is viable in GRB environments.  In any
case, the origin of such clouds is a genuine difficulty which needs to
be addressed.  

There are additional concerns regarding the geometry of the cloud
region.  From eq.~\ref{peak_fluence}, we see that the magnitude of the
fluence at the spectral peak behaves as $dE/dA \propto R^{-4}$ for
clouds of the same density and size.  Therefore, in order for the
pulses to have roughly equal fluence throughout the burst light curve,
the clouds must either be confined to a narrow range in radius and
essentially form a thin shell of clouds, or the cloud density and size
distributions must compensate for the strong decline in fluence at
larger radii.  If the clouds are confined to a thin shell, then for
the parameters we have considered, only $\sim 25$ clouds would be
required to cover the shell entirely within the initial Doppler
beaming cone of the blast wave.  Constraints such as this may place
severe limits on the amount of variability which can be obtained in
this model.

The microphysics of the relativistic shock also needs to be more
carefully considered in all synchrotron shock models.  We have assumed
that the magnetic field energy density will always be a constant
fraction of the energy density of the shocked material.  There is no
particular reason that this will be the case.  Lastly, we have allowed
the particle distributions to accumulate without regard to their
effect on the state of the shocked fluid.  If it is not radiated, some
of this particle energy will result in the expansion of the shock
region and there will then be adiabatic losses to consider.

Despite the numerous simplifications of this model of blast wave/cloud
interactions, the central point is still quite valid: if synchrotron
radiation from relativistic shocks is responsible for the prompt
emission of GRBs, then the bulk deceleration of the shocked material
can have a profound effect on the observed time-integrated spectra.
Therefore any conclusions regarding the radiative efficiency of these
shocks based upon spectral shapes must also consider these effects.
More optimistically, if this model is relevant to burst temporal
variability, then calculations such as this can provide a starting
point for using GRB light curves as a probe of the circumburst
environment and may then provide some insight into the actual
progenitors of these events.

\acknowledgements
The author acknowledges helpful discussions with C. D. Dermer and
E. E. Fenimore.  I also thank M. A. Nowak for useful comments which
have improved the text.  Part of this work was performed while the
author held a National Research Council-NRL Research Associateship.


\begin{thebibliography}{999}
\bibitem{C98} Chiang, J. 1998, ApJ, 508, in press (astro-ph/980530)
\bibitem{CD98} Chiang, J., \& Dermer, C. D. 1998, ApJ, in press 
   (astro-ph/9803339)
\bibitem{C97} Cohen, E., et al. 1997, ApJ, 488, 330
\bibitem{D97} Dermer, C. D., Sturner, S. J., \& Schlickeiser, R. 1997,
   ApJS, 109, 103
\bibitem{D98} Dermer, C. D. 1998, ApJ, 501, L157
\bibitem{DCB98} Dermer, C. D., Chiang, J., \& B\"ottcher, M., 1998, ApJ,
   in press (astro-ph/9804174)
\bibitem{DM98} Dermer, C. D., \& Mitman, K. E. 1998, ApJL, submitted 
   (astro-ph/9809411)
\bibitem{F95} Fenimore, E. E., et al. 1995, ApJ, 448, L101
\bibitem{F96} Fenimore, E. E., Madras, C. D., \& Nayakshin, S. 1996,
   ApJ, 473, 998
\bibitem{F98} Fenimore, E. E., et al. 1998, ApJ, submitted 
   (astro-ph/9802200)
\bibitem{F97} Frail, D., et al. 1997, Nature, 389, 261
\bibitem{K94} Katz, J. I. 1994, ApJ, 432, L107
\bibitem{KTH98} Kazanas, D., Titarchuk, L. G., \& Hua, X.-M. 1998, ApJ, 
   493, 708
\bibitem{K98} Kulkarni, S. R., et al. 1998, Nature, 393, 35
\bibitem{M97} Metzger, M. R., et al. 1997, Nature, 387, 878
\bibitem{N96} Norris, J., et al. 1996, ApJ, 459, 393
\bibitem{P98} Preece, R. D., et al. 1998, ApJL, in press (astro-ph/9808184)
\bibitem{RL79} Rybicki, G. B., \& Lightman, A. P. 1979, Radiative
   Processes in Astrophysics (New York: Wiley)
\bibitem{SP95} Sari, R., \& Piran, T. 1995, ApJ, 455, L143
\bibitem{SP97} Sari, R., \& Piran, T. 1997, ApJ, 485, 270
\bibitem{SPN98} Sari, R., Piran, T., \& Narayan, R. 1998, 497, L17
\bibitem{S98} Schaefer, B. E., et al. 1998, ApJ, 492, 696
\bibitem{T96} Tavani, M. 1996, ApJ, 466, 768
\end{thebibliography}
\end{document}